\documentstyle[twoside,12pt,epsf]{article}
\renewcommand{\baselinestretch}{1.2}

\begin{document}

\newcommand{\ch}{\raisebox{0.5mm}{$\chi$}}
\newcommand{\D}{\displaystyle}
\newcommand{\dec}{\rightarrow}
\newcommand{\epr}{\eta^{\prime}}
\newcommand{\ew}[1]{\left\langle #1 \right\rangle}
\newcommand{\Kdelta}{\raisebox{0.5mm}{$\delta$}}
\newcommand{\ket}[1]{| #1 \,\rangle}
\newcommand{\kkq}{K\bar{K}}
\newcommand{\kmin}{K^-}
\newcommand{\kmins}{K^{-*}}
\newcommand{\knq}{\bar{K}^0}
\newcommand{\knqs}{\bar{K}^{0*}}
\newcommand{\knull}{K^0}
\newcommand{\knulls}{K^{0*}}
\newcommand{\kpl}{K^+}
\newcommand{\kpls}{K^{+*}}
\newcommand{\op}[1]{\hat{#1}}
\newcommand{\Pj}{{\cal P}}
\newcommand{\pmin}{\pi^-}
\newcommand{\pnull}{\pi^0}
\newcommand{\ppl}{\pi^+}
\newcommand{\pqp}{\bar{p}p}
\newcommand{\rnull}{\rho^0}
\newcommand{\SS}{\scriptstyle}
\newcommand{\SSS}{\scriptscriptstyle}
\newcommand{\tr}{{\rm tr}}
\newcommand{\TS}{\textstyle}
\newcommand{\T}{\textstyle}
\newcommand{\lsim}{\raisebox{-0.8ex}{$\stackrel{\textstyle<}{\sim}$}}
\newcommand{\gsim}{\raisebox{-0.8ex}{$\stackrel{\textstyle>}{\sim}$}}

\renewcommand{\thefootnote}{\fnsymbol{footnote}}

\begin{titlepage}
 \vspace*{-1cm}

 \setlength{\baselineskip}{25pt}
 \centerline{{\huge U}\large niversit\"at {\huge R}egensburg}
 \centerline{\large Institut f\"ur Theoretische Physik}

 \begin{center}
 \begin{minipage}[t]{7cm}
 \epsfysize=5cm\epsfbox{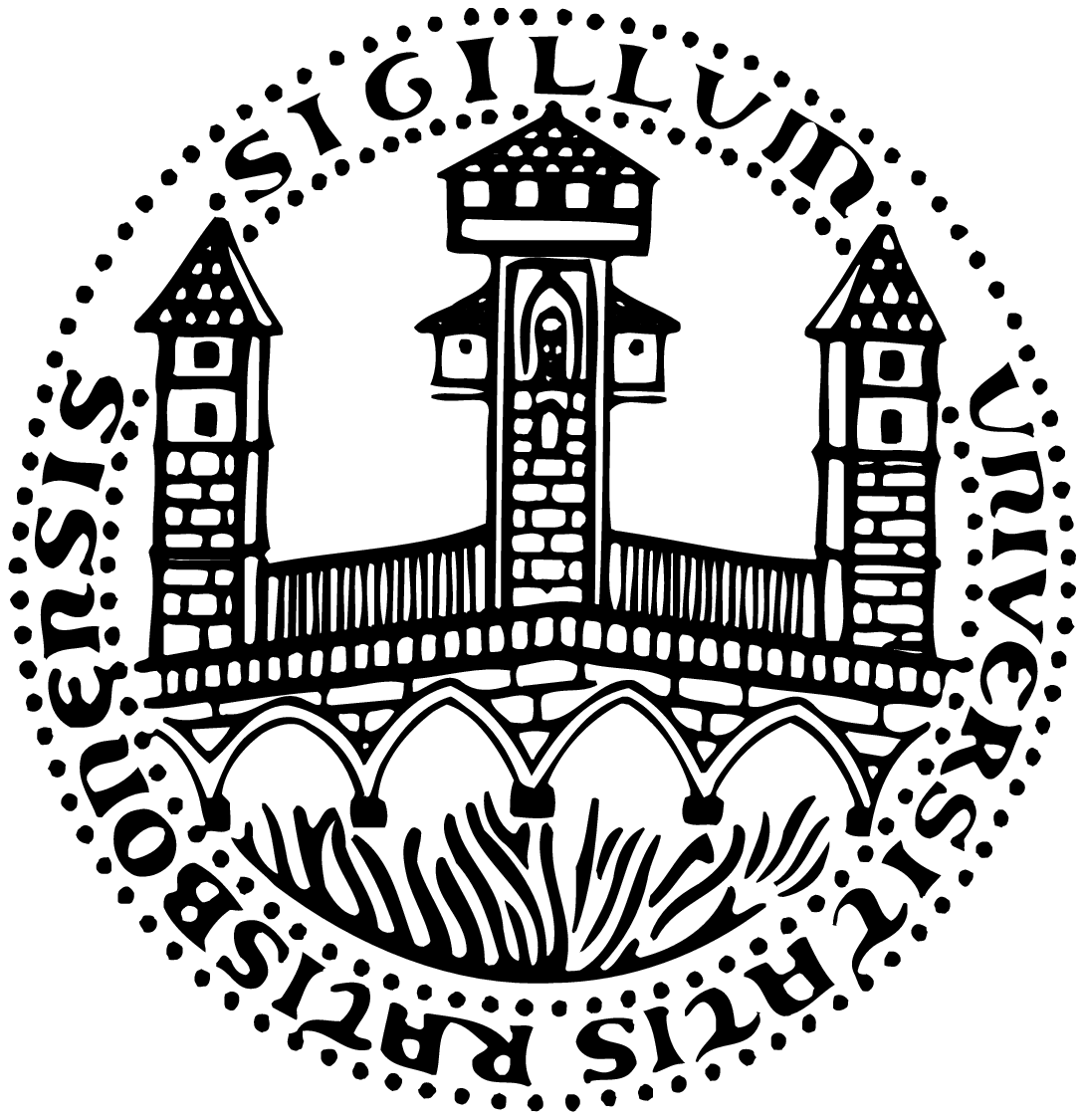}
 \end{minipage}
 \end{center}

 \vspace{1cm}

 \setlength{\baselineskip}{25pt}
 \centerline{\Large PION MULTIPLICITY DISTRIBUTION IN }
 \centerline{\Large ANTIPROTON-PROTON ANNIHILATION AT REST\footnote{Work
 supported in part by DFG, BMFT and GSI}}

 \vspace{0.5cm}

 \small\normalsize
 \centerline{Wolfgang Bl\"umel and Ulrich Heinz}

 \vspace{0.3cm}

 \centerline{\small Institut f\"ur Theoretische Physik, Universit\"at
 Regensburg,}
 \centerline{D-93040 Regensburg, Germany} 

 \vskip 1.0cm

 \noindent
 \renewcommand{\baselinestretch}{1.0}
 \small\normalsize
 {\bf Abstract.}
 The pion multiplicity distribution is widely believed to reflect the
 statistical aspects of $\bar{p}p$ annihilation at rest. We
 try to reproduce it in a grand canonical picture with explicit conservation of
 electric charge, isospin, total angular momentum, and the parity
 quantum numbers $P$, $C$, and $G$ via the projection operator formalism.
 Bose statistics is found 
 to be non-negligible, particularly in fixing the interaction volume.
 The calculated pion multiplicity distribution for $\left\langle 
 n_{\pi} \right\rangle = 5$ 
 turns out to depend strongly on the conservation of the angular
 momentum and connected quantum numbers, as well as on the 
 spin state occupation in S-wave annihilation. However, the empirical
 Gaussian pion multiplicity distribution 
 cannot be reproduced. This calls in question either the statistical ansatz
 or the rather old data themselves. 

 \vfill

 \centerline{\hfill 09/13/94\ \ TPR-94-30}

\end{titlepage}
%
%
%%%%%%%%%%%%%%%%%%%%%%%%%%%%%%%%%%%%%%%%%%%%%%%%%%%%%%%%
\section{Introduction}
%%%%%%%%%%%%%%%%%%%%%%%%%%%%%%%%%%%%%%%%%%%%%%%%%%%%%%%%
%
%
Antinucleon-nucleon annihilation is dominated by nonperturbative QCD
effects, and up to now there exists no consistent description of
its microscopic dynamics within QCD due to the complexity of the
underlying mechanisms. It turns out, however, that a number of
global features can be reproduced at least qualitatively by 
phenomenological approaches (for a review, see \cite{Dov,Ams}). 
In particular, statistical models in several variations have been
applied to the $\pqp$ annihilation system (for a review of the early
works, see \cite{Kre}), and even the fireball picture has been
used motivated by the Maxwell-Boltzmann like form of the pion-energy
spectra \cite{Ghe}--\cite{Kim}.

\begin{figure}[hbt]
 \begin{minipage}[t]{9cm}
 \makebox[0cm]{}\\
 \epsfysize=7cm\epsfbox{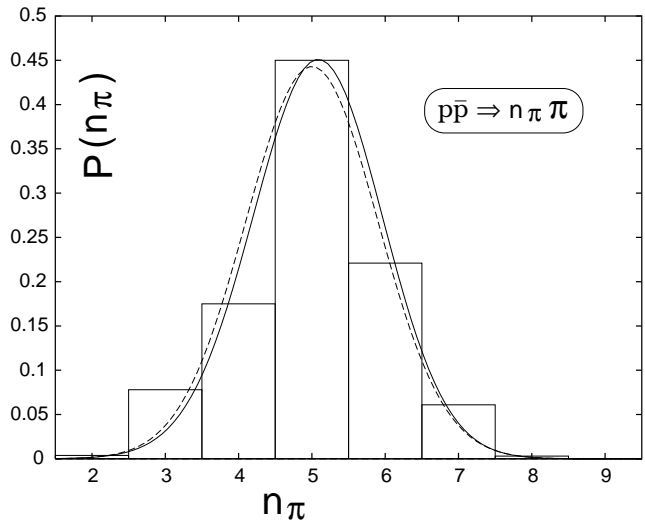}
 \end{minipage}
 \hfill 
 \parbox[t]{6cm}{
 \caption{\label{emp} \sl Pion distribution of $\pqp$ annihilation 
 at rest. The empirical data (histogram) [4] are compared with a Gaussian
 distribution (full line) and a modified Poissonian (dashed line). }}
\end{figure}

Especially the pion multiplicity distribution has always been
interpreted as a strong hint for a statistical behaviour of 
the $\pqp$ annihilation system: the data \cite{Ghe}
seem to be fitted by a Gaussian distribution 
\begin{equation} \label{gauss}
 P(n_{\pi}) = \frac{1}{\sqrt{2\pi}D}\exp\left[ 
 -\frac{(n_{\pi} - \ew{n_{\pi}})^2}{2D^2}\right]\:\:,
\end{equation}	
with $D = 0.9$ and $\ew{n_{\pi}}= 5$.
Fig. \ref{emp} also shows in addition a modified
Poissonian distribution, which fits the data equally well,
\begin{equation} \label{Poisson}
 P(n_{\pi}) = C\cdot \frac{\lambda^{n^a}\cdot\exp(-\lambda)}
                          {(n^a)!}\:\:,
\end{equation}
where $\lambda = 14$ and $a = 1.6$. It is therefore natural to
try to reproduce the empirical pion distribution by using
a statistical model.

On the other hand, conservation of the characteristic quantum numbers
of the $\pqp$ system, like isospin \cite{Mag}--\cite{Mul} or total angular
momentum \cite{May1}--\cite{Blu}, are known to play an important role
and to severely constrain the statistical phase space. In this paper we
address the question how these conservation laws influence the pion
multiplicity distribution in $\pqp$ annihilation at rest. 
Comparison of our calculations with the data is hampered by a serious
experimental problem:
the occupation of the various spin-states $J^{PC}$ of the protonium atom just 
before its annihilation is apparently not distributed in a statistical way,
but depends on the density of the hydrogen target and cannot be 
experimentally controlled up to now \cite{Rot,Gas}.
\small
\normalsize
S-wave annihilation occurs from either a spin singlet $(J=0)$ or a 
spin triplet $(J=1)$ state, but with unknown relative occupation. 
Therefore, we here study systematically the gross features of S-wave
$\pqp$ annihilation (e.g. the pion multiplicity distribution) as a 
function of the ratio
 \[
 X_r = \frac{W(0^{-+})}{W(1^{--})}\qquad \bigg({\rm where}\quad 
 W(0^{-+}) + W(1^{--}) = 1\bigg)\:,
 \]
between the spin singlet and triplet occupation probabilities.

In our approach, we will use a statistical model in connection with 
the projection operator formalism which allows for the explicit conservation
of isospin as well as of the so-called external quantum numbers, i.e. the
angular momentum and $P$, $C$, and $G$ parities. Instead of the usual
Boltzmann approximation, Bose statistics is employed 
(for reasons that will be discussed) and
all mesons and resonances up to the $a_1(1260)$ are included in our 
calculations. At a fixed average pion multiplicity of $\ew{n_{\pi}} = 5$ we 
discuss the dependence of the volume parameter on temperature and
then present the pion multiplicity distributions for several values of $X_r$.
%
%
%
%%%%%%%%%%%%%%%%%%%%%%%%%%%%%%%%%%%%%%%%%%%%%%%%%%%%%%%%
\section{The Model}
%%%%%%%%%%%%%%%%%%%%%%%%%%%%%%%%%%%%%%%%%%%%%%%%%%%%%%%%
%
%
It is rather easy to implement baryon number and strangeness
conservation into the statistical approach: one builds the total partition
function from multi-mesonic states only, each of which 
has vanishing strangeness $S=0$.
Conservation of the electric charge $Q$ is included automatically once 
the third component of the isospin is conserved, i.e. $I_3 = 0$,
and the constraining effects of $P,C$, and $G$ parity conservation
can easily be taken into account \cite{Blu} if the angluar momentum is fixed,
which together with the total spin is subject to
the conservation of the total angular momentum of the system.
So, it will be sufficient to set up a procedure which allows for the
projection of the total partition function onto the isospin quantum numbers
$(I,I_3)$ and the angular momentum $L$.

A convenient and, above all, consistent way of treating 
conservation of both Abelian and non-Abelian quantum numbers in the
context of a statistical model is provided by the projection operator 
formalism. The simplest procedure would be to calculate the grand canonical 
partition function $Z_{gc}$ of the produced mesons, which are supposed to 
behave like free particles in a spherical ``box'' while the interaction
between these mesons is accounted for by resonance production \cite{Bel,Ser}.
The unphysical multi-meson states which are contained in $Z_{gc}$
due to quantum number fluctuations then have to be projected out by 
suitable projection operators.
This is a well established method {\cite{Red} -- \cite{Elz}}
which has been mostly applied to
internal quantum numbers like $U(1)$-baryon number or -strangeness
and $SU(2)$-isospin, but can also be extended to treat ``external'' quantum 
number conservation like that of angular momentum \cite{Blu}.
Following this line, the trick is not to calculate $Z_{gc}$ itself
but a so-called generating function $\tilde{Z}$. In the case of isospin and
angular momentum conservation (AMC), the generating function can be
written as
 \begin{equation} \label{mark1}	
	\tilde{Z}(T,V,\vec{\alpha}_I,\vec{\alpha}_L) = 
	\tr\,[\exp(-\beta\hat{H} + i\vec{\alpha}_I\cdot\vec{I} +
                                      i\vec{\alpha}_L\cdot\vec{L})]\:,
 \end{equation}			
with the symmetry group parameters $\vec{\alpha}_I$ and
$\vec{\alpha}_L$ for the $SU(2)$-isospin and $SO(3)$-rotation group,
respectively. It can be easily shown, that this generating function can 
be expressed in an alternative way, namely
 \begin{equation}
	\tilde{Z}(T,V,\vec{\alpha}_I,\vec{\alpha}_L) =
	\sum_{I,L}\frac{Z_{I,L}(T,V)}{(2I+1)(2L+1)}\cdot
	\ch_I(\vec{\alpha}_I)\cdot\ch_L(\vec{\alpha}_L)\:.
 \end{equation}
Here, $\ch_I$ and $\ch_L$ are the group characters \cite{Gre}, and the 
summation
has to be done over all irreducible representations of the symmetry
groups under consideration. $Z_{I,L}$ is a constrained partition
function which contains only those multi-meson states which are eigenstates
of $\hat{\bf I}^2$ and $\hat{\bf L}^2$ with eigenvalues $I(I+1)$ and
$L(L+1)$ and which is the actual quantity in
demand. In order to project $\tilde{Z}$ on $Z_{I,L}$, the orthogonality
relation of the group characters has to be wrapped up into a
corresponding projection operator. 

Let us now write down some explicit expressions for the general case
of $r$ different particle species being indexed by 
$j = 1,\ldots,r$. In particular, the particle species are
distinguished by their isospin $t$, third component $t_3$ and mass, i.e.
the index $j\,$ is only a short notation for $(t,t_3,m)$.
Furthermore, two particles are considered to be identical if they 
correspond in $j,l$ and their one particle energy $\epsilon(k)$.
Because the geometry of the quantum gas resulting from $\pqp$ annihilation 
at rest is supposed to be approximately spherically symmetric, the 
projection onto angular momentum eigenstates should not prefer a certain 
direction. Hence, no projection on the third component of the angular 
momentum is needed, and for this reason we are allowed to replace  
$\hat{\bf L}$ by $\hat{L}_3$ \cite{Gre} in (\ref{mark1}).
In order to get the generating function of the system,
the trace in equation (\ref{mark1}) is worked out with 
eigenstates of $(\hat{\bf I}^2, \hat{I}_3)$ and $(\hat{\bf L}^2,\hat{L}_3)$, 
and the occupation number representation is employed (for details, see 
\cite{Blu}): 
 \begin{eqnarray}
    \tilde{Z}(T,V,\vec{\alpha},\omega) & = &
    \tr \left[ \exp({ -\beta\hat{H} + i\vec{\alpha}\vec{I} 
                                 + i\omega\hat{L}_3 } )  \right]
     \nonumber \\[1ex]
    &  & \hspace{-2cm} 
            =   \prod_{j=1}^r
	        \prod_{l=0}^{\infty}
		\prod_{k_l=0}^{\infty}   
    \left[ 
           \Big( D_{t_3^j,t_3^j}^{t^j}(\vec\alpha) \Big)^{n_{l,k_l}^j}
	   \cdot \Big(\ch^l(\omega)\Big)^{n_{l,k_l}^j}\cdot
           \exp\Big( n_{l,k_l}^j(-\beta\epsilon_{k_l}^j )\Big) 
    \right] \:\:, \label{mark2}
 \end{eqnarray}
where $\beta = 1/T$ and $\epsilon_{k_l}^j$ is the relativistic energy of
a particle of type $j$. The occupation numbers
$n_{l,k_l}^j$ denote the number of particles of type $j$ in a
state with angular momentum $l$ and magnitude
of momentum $k_l$. Note, that the momentum $k$ depends on $l$ and the
interaction volume $V$, because the
radial component of the angular momentum eigenfunction is 
a spherical bessel function, $j_l(kr)$, which must vanish at the edge of
the spherical ``box''.

Usually, at this stage the Boltzmann approximation is invoked. The
argument is \cite{Mul,Koc} that Bose statistics makes at most only a ten 
percent
correction in the case of the one pion partition function, and the correction
decreases rapidly with increasing particle mass. However, we will show
that when going to multi-mesonic states then the error resulting from the
Boltzmann approximation can be much larger and Bose statistics must be taken
into account. Thus, after some simple transformations
and proceeding with Bose statistics, equation (\ref{mark2}) can be rewritten
in the following form:
 \begin{equation}  \label{mark3}
    \tilde{Z}(T,V,\vec{\alpha},\omega) =
	\exp \left[ \sum_{n=1}^{\infty}\sum_{j=1}^r
	\frac{1}{n} \Big( D_{t_3^j,t_3^j}^{t^j}(\vec\alpha) \Big)^n\cdot 
	\sum_{l=0}^{\infty} z_{n,l}^j(T,V)\cdot
	\Big(\ch^l(\omega)\Big)^n  \right]
 \end{equation}
Due to Bose statistics, more than one particle of the same type can 
occupy a single level characterized by its angular
momentum eigenvalue $l$ and its energy $\epsilon(k_l)$. In the above
expression, the partition function for $n$ particles of type $j$, each with
the same angular momentum $l$, is given by
 \begin{equation} \label{mark4}
    z_{n,l}^j(T,V) \equiv \sum_{k_l = 0}^{\infty} 
    \exp (-\beta\epsilon_{k_l}^j\cdot n) \:\:,
 \end{equation}
where the sum is over all energy eigenvalues for the given value of $l$
(obtained as zeroes of the spherical Bessel function, see above).
The classical Boltzmann approximation is contained in (\ref{mark3})
by considering only the term $n=1$.

In general, the form of the projection operators depends on the choice of the
symmetry group parameters $\vec{\alpha}_I$ and $\vec{\alpha}_L$. 
For convenience we choose for $SU(2)$-isospin 
the Euler angles $(\alpha,\beta,\gamma)$ and for $SO(3)$-angular
momentum the unit vector $\vec{n}$ of the rotation axis together with 
the rotation angle $\omega$. Then, we can write for the isospin
projector
 \begin{equation}
   \hat{P}_{\SSS I,I_3=0} = \frac{2I+1}{8\pi^2}
                      \int\limits_0^{2\pi}d\alpha
                      \int\limits_0^{\pi}d\beta
                      \int\limits_0^{2\pi}d\gamma\,  
         \sin\beta\, P_{\SSS I}(\cos\beta)\:\:,
 \end{equation}
where the group character $\ch_I$ was replaced by the matrix element
$D_{0,0}^I(\alpha,\beta,\gamma) = P_I(\cos(\beta)$,
because we want to project onto $(I,I_3=0)$ rather than only $I$. 
In the case of AMC we get
 \begin{equation}
    \hat{P}_{\SSS L} = \frac{2L+1}{2\pi^2}
                       \int d\vec{n}\int\limits_0^\pi d\omega 
        \sin\left({\TS\frac{\omega}{2}}\right)
        \sin\left( (2L+1){\TS\frac{\omega}{2}}\right)
        \:\:.
 \end{equation}
The only way to apply these projectors directly to $\tilde{Z}$ of equation
(\ref{mark3}) is to expand the exponential and the resulting expressions
until we have a sum over all possible final multi-meson states contained
in the generating function. Because of the huge amount of such 
states (all mesons up to the $a_1(1260)$ are included in the calculations),
the evaluation is done by the computer. In particular the 
partition functions $z_{n,l}^j(T,V)$ are evaluated numerically.
Furthermore, the evaluation remains treatable only up to a total number
of $N=7$ mesons. This cut-off in the partition function can be well
justified by looking at the data: the contribution of channels with
$N > 7$ is negligible.
%
%
%
%%%%%%%%%%%%%%%%%%%%%%%%%%%%%%%%%%%%%%%%%%%%%%%%%%%%%%%%%%%%%%
\section{Average Pion Mulitiplicity and Multiplicity Distribution}
%%%%%%%%%%%%%%%%%%%%%%%%%%%%%%%%%%%%%%%%%%%%%%%%%%%%%%%%%%%%%%
%
%
In a first step, we want to adjust the model parameters $T$ and $V$
to the empirical pion expectation value $\ew{n_\pi} = 5$. 
In addition to the multiplicity of the directly produced pions, one
must also calculate the multiplicities of all the resonances in the
system which then decay according to \cite{Par} into ``secondary'' pions. 

In order to extract multiplicities from the total partition function
of the system $Z(T,V,B,S,I,\ldots)$,
we introduce for every particle type $j$ a chemical
potential $\mu_j$, and every partition function $z_{n,l}^j$ has to be 
multiplied by the factor 
 \begin{equation}
      (\lambda^j)^n = \exp{\beta\mu_j\cdot n}\:\:.
 \end{equation} 
Then, the particle mulitplicities result from the derivative 
of the logarithm of the total partition function with respect to the
fugacities at $\lambda^j = 1$,
 \begin{equation}
   \ew{N_j} = \partial_{\lambda^j}\ln\left[ 
             Z(T,V,B,S,I,\ldots,\lambda^j,\ldots) \right]_{\lambda^j=1}\:\:.
 \end{equation}

\begin{figure}[hbt]
 \begin{minipage}[t]{8cm}
 \makebox[0cm]{}\\
 \epsfysize=8cm\epsfbox{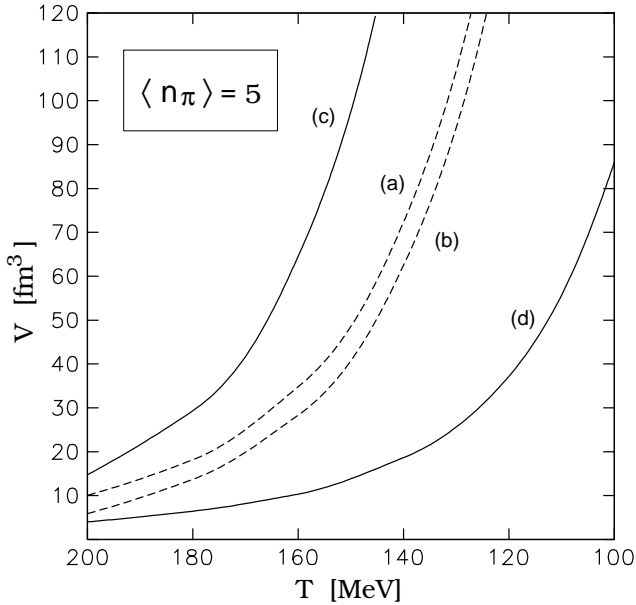}
 \end{minipage}
 \hfill 
 \parbox[t]{7cm}{
 \caption{\label{vtplot}\sl Dependence of the interaction volume $V$ on  
 temperature $T$ for $\ew{n_{\pi}} = 5$ for various scenarios: {\bf (a)} 
 Boltzmann statistics, isospin conservation only; {\bf (b)} Bose statistics, 
 isospin conservation only; {\bf (c)} Boltzmann statistics, additional AMC 
 and conservation of $C$, $P$, and $G$ parity; {\bf (d)} Bose statistics, 
 additional AMC and conservation of $C$, $P$, and $G$ parity.}
 }
\end{figure}

In Fig.~\ref{vtplot} the interaction volume of the $\pqp$ annihilation
system is plotted against the temperature of the system for a fixed value
of $\ew{n_{\pi}} = 5$. The dashed lines correspond to $(I,I_3)$ conservation
only. Here, the deviation of Boltzmann statistics in 
curve (a) from Bose statistics (b) is not
very significant. However, if AMC and the conservation of
the parity quantum numbers are included in our calculations, then
the difference is striking: Bose statistics (d)
allows for a rather small and therefore much more realistic volume in a 
temperature range of $T=[140\ldots 200]\ \rm{MeV}$ whereas the Boltzmann case
(c) yields a reasonable volume only for very high temperatures 
of $T\ \gsim\ 200\ \rm{MeV}$. This is a significant phenomenological
improvement compared to previous studies which were plagued by
unrealistically large fireball volumes \cite{Kre,Mul}.
\small
\normalsize
This effect is obviously an intricate combination of both Bose statistics 
{\sl and} conservation of external quantum numbers: the Boltzmann
curves (a) and (c) show opposite behaviour, compared to the Bose curves
(b) and (d), when going from the dashed to the full lines, i.e when
including AMC and conservation of the parity quantum numbers. 
Note, that in (c) and (d) we did not distinguish between the 
protonium spin states, because in our calculations the volume is nearly
the same (up to 5 percent) for $0^{-+}$ and $1^{--}$. This means, that
in our model the volume is independent of the spin-state ratio $X_r$.

\begin{figure}
 \begin{minipage}{16cm}
 \epsfysize=7cm\epsfbox{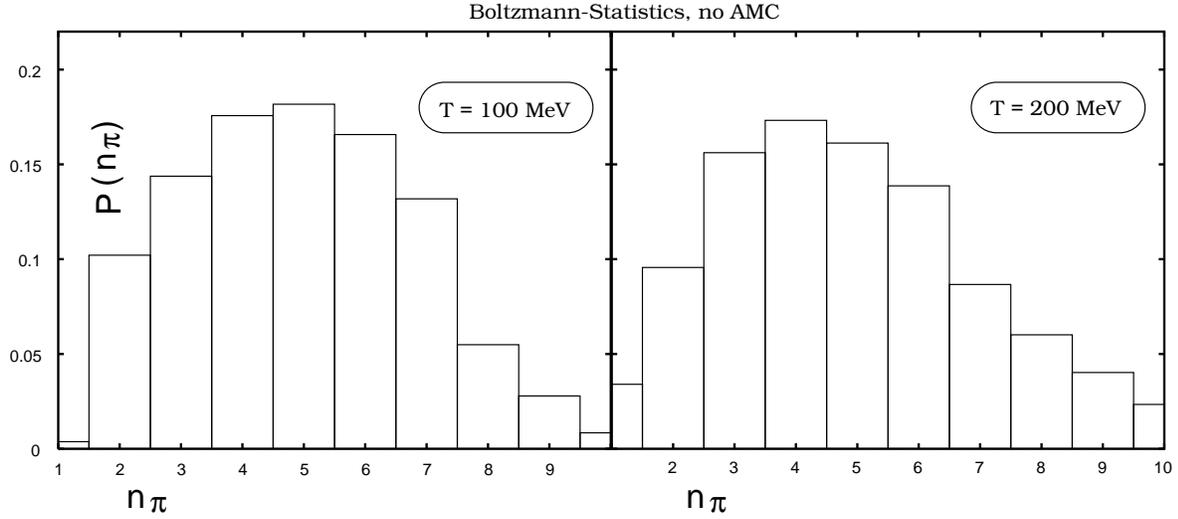}
 \end{minipage}
 \caption{\label{mbaus}\sl Pion multiplicity  distribution with 
 $\ew{n_{\pi}} = 5$, for $T = 100\ MeV$ and $T = 200\ MeV$.
 The Boltzmann approximation is used and only the isospin and its third
 component are conserved.}
\end{figure}

Once the parameters $T$ and $V$ are known, the pion multiplicity 
distribution can be calculated in a second step. To this end we use the
following trick: After expanding 
the projected partition function into a sum over all the allowed
multi-meson channels, we multiply each channel by 
\begin{equation} \label{mark5}
 1 = \sum_{n_\pi = 0}^\infty p(n_{\pi})\:\:,
\end{equation}
where $p(n_{\pi})$ is the probability (calculated from the data tables
\cite{Par}) that, after all resonances have decayed, this particular
channel yields $n_\pi$ final pions. Then, in order to be able to
project out a particular final state with $n_\pi$ pions, we multiply each
term in the sum on the right hand side of eq. (\ref{mark5}) with a 
Lagrange multiplier $f(n_\pi)$. Explicitly, we make the replacement
 \begin{equation}
    z_{n_1,l_1}^{j_1}\cdots z_{n_r,l_r}^{j_r} \longmapsto
    \sum_{n_\pi = 0}^\infty z_{n_1,l_1}^{j_1}\cdots z_{n_r,l_r}^{j_r}
    \cdot p_{n_1,\ldots,n_r}^{j_1,\ldots,j_r}(n_\pi)\cdot f(n_\pi)\:\:,
 \end{equation}
where $p_{n_1,\ldots,n_r}^{j_1,\ldots,j_r}(n_\pi)$ is the probability
for having $n_\pi$ final pions produced by a channel containing $n_1$
resonances of type $j_1$, $n_2$ resonances of type $j_2$, etc.
If the logarithm of the so prepared partition function is 
derived with respect to $f(n_\pi)$ (setting $f(n_\pi) = 1$ afterwards), 
we get the total probability of having $n_\pi$ pions in the final state
as
 \begin{equation}
	P(n_\pi) = \partial_{f(n_\pi)} \ln \Big[ 
	Z(T,V,B,S,I,\ldots,f(n_\pi),\ldots) \Big]_{f(n_\pi) = 1}\:\:.
 \end{equation}
The calculated pion distribution for the simplest case of isospin
conservation only in the Boltzmann approximation is shown in 
Fig.~\ref{mbaus} for two extreme temperatures, $T = 100\ \rm{MeV}$ and
$T = 200\ \rm{MeV}$.
\small
\normalsize
\small
\begin{figure}
 \begin{minipage}{16cm}
 \epsfysize=7cm\epsfbox{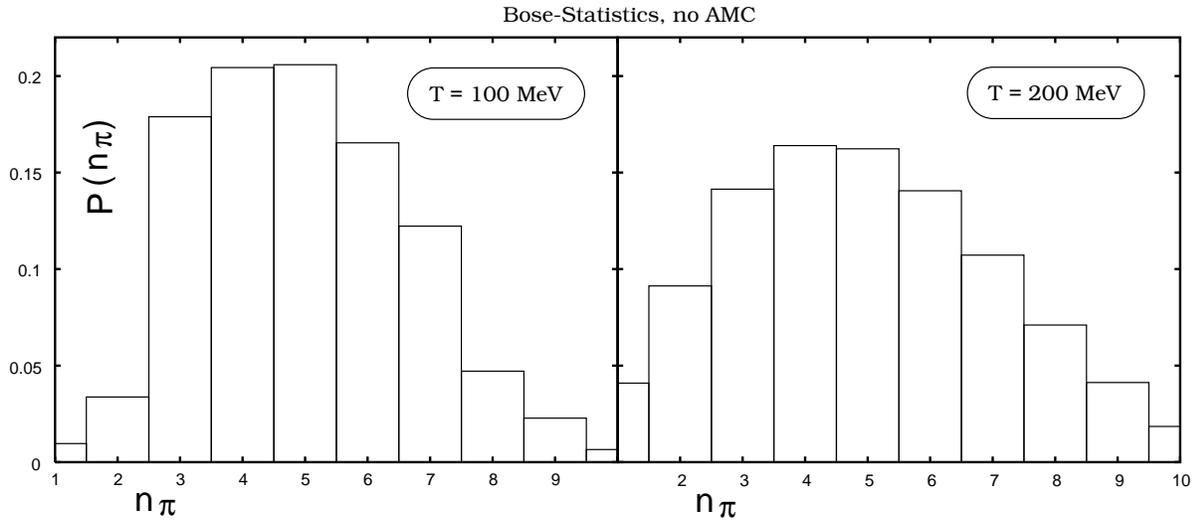}
 \end{minipage}
 \caption{\label{beaus}\sl Pion multiplicity distribution with 
 $\ew{n_{\pi}} = 5$, for $T = 100\ MeV$ and $T = 200\ MeV$ with
 Bose statistics.  Only isospin $I$ and its third component $I_3$ are
 conserved.}
\end{figure} 
\normalsize
The main effect of high temperatures is to give the high mass resonances
more weight in the partition function, and the resulting pion distribution
should be broader than at low temperatures. Especially for Boltzmann 
statistics,
however, this effect seems to be rather small, see Figure \ref{mbaus}.
The Bose case is a little bit more sensitive to the temperature, although
the effect is also not very drastic (Fig.~\ref{beaus}). The shapes of both
distributions resemble that of a Poissonian, as it should be in a grand 
canonical ensemble \cite{Cle}, but this Poissonian is slightly modified
by resonance decays and isospin conservation. 
Compared with the empirical distribution of Fig.~\ref{emp}, all the
curves of Figs. \ref{mbaus} and \ref{beaus} are much too broad. This
means that the constraint on the partition function originating from
isospin conservation alone is too weak, and that there are still too many
open channels broadening the pion distribution. 

\begin{figure}
 \begin{center}	
 \begin{minipage}{16cm}
 \epsfysize=18.5cm\epsfbox{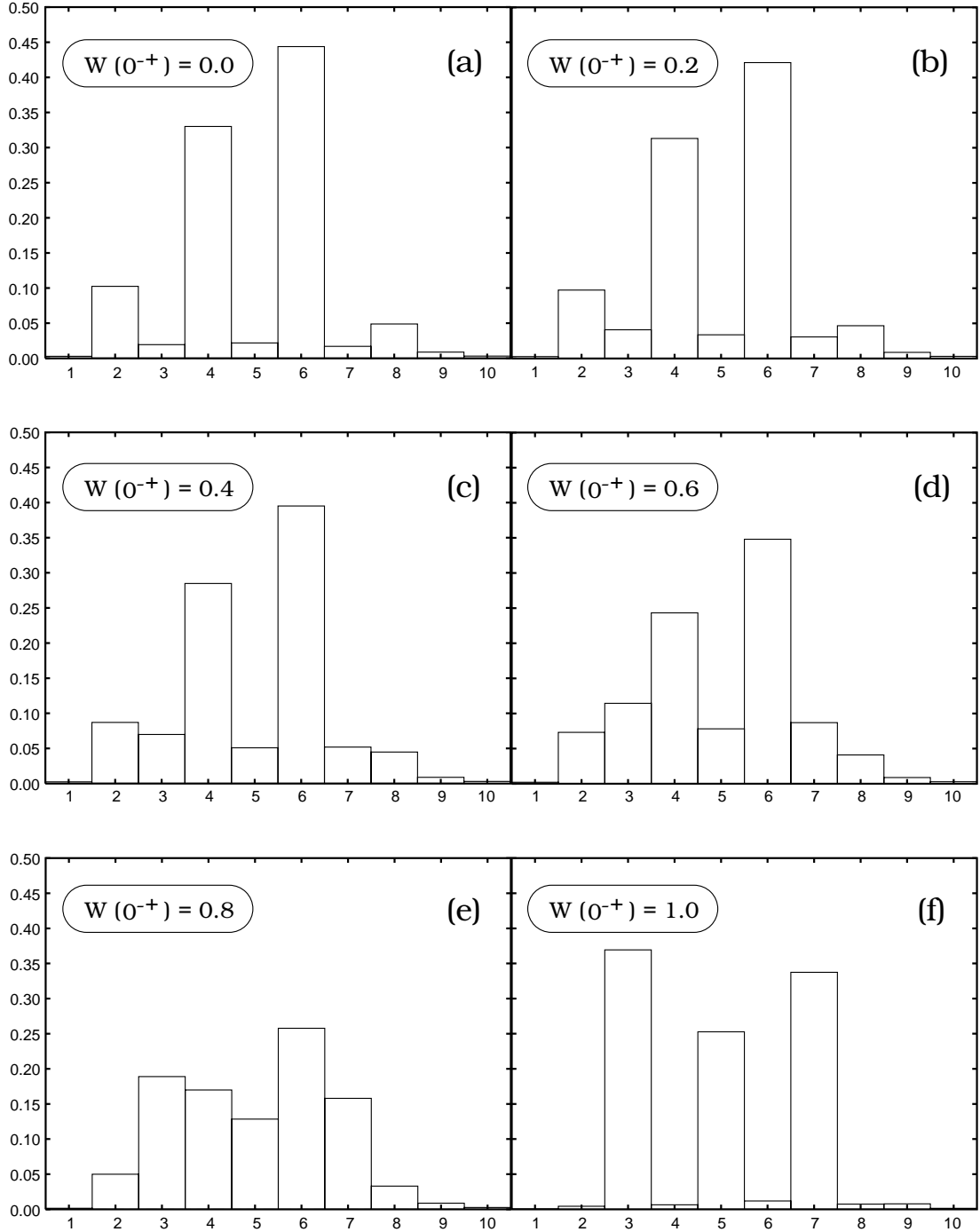}
 \end{minipage}
 \end{center}
 \caption{\label{beein}\sl Pion multiplicity distributions with 
 $\ew{n_{\pi}} = 5$, for $T = 160\ MeV$ using
 Bose statistics. In addition to the isospin conservation, also
 AMC and conservation of $P$, $C$, and $G$ parity are included.
 Six different values for the spin-state occupation of the
 protonium are taken into account, which are characterized by
 {\bf (a)} $W(0^{-+}) = 0.0$, {\bf (b)} $W(0^{-+}) = 0.2$,
 {\bf (c)} $W(0^{-+}) = 0.4$, {\bf (d)} $W(0^{-+}) = 0.6$,
 {\bf (e)} $W(0^{-+}) = 0.8$, and {\bf (f)} $W(0^{-+}) = 1.0$.
 }
\end{figure}

Now, by additional conservation of ``external'' quantum numbers a
stronger constraint is put on the number of possible multi-meson channels.
Due to the weak temperature dependence of the distribution and the
large numerical effort, we have restricted
the calculations to a single temperature value of $T = 160\ \rm{MeV}$. 
In order to account for the unknown ratio $X_r$ of the spin-state occupation
of protonium just before annihilation, we have plotted the pion
distribution for several values of $W(0^{-+})$ in Fig.~\ref{beein},
using Bose statistics.
\small
\normalsize
The behaviour of the calculated pion distribution with increasing
$W(0^{-+})$ can be summarized as follows. Up to a value of
$W(0^{-+}) = 0.6$ (Figs. \ref{beein}a to \ref{beein}d), the 
{\it even} numbered $n_{\pi}$ states dominate strongly, especially the
$n_{\pi} = 4$ and $n_{\pi} = 6$ states. For $W(0^{-+}) = 0.8$ 
(Fig.~\ref{beein}e) the
states with $n_{\pi} = 3$ up to $n_{\pi} = 7$ are rather equally
distributed. Finally, in the limit of exclusive occupation 
of the $0^{-+}$
state, the {\it odd} numbered states with $n_{\pi} = 3,5,$ and $7$ dominate
the distribution, see Figure~\ref{beein}f. For statistical
occupation of the spin states according to their combinatoric weight,
$W(0^{-+}) = 0.25$, we would have a result close to that of Fig.~\ref{beein}b.
All in all, the obvious conclusion is that generally it is very difficult
to get a Poissonian or even a Gaussian distribution when
the external quantum numbers are conserved. This complete destruction of
a smooth distribution by the conservation of external quantum
numbers is clearly a consequence of the large fraction of
directly produced pions in the partition function, which overwhelms the
indirect production through resonance decays due to the larger phase
space. 
We want to emphasize here that this is not an artefact
of our grand canconical approach, but should also be observed in 
microcanonical calculations.
%
%
%
%%%%%%%%%%%%%%%%%%%%%%%%%%%%%%%%%%%%%%%%%%%%%%%%%%%%%%%%%%%%%%%%%
\section{Conclusions}
%%%%%%%%%%%%%%%%%%%%%%%%%%%%%%%%%%%%%%%%%%%%%%%%%%%%%%%%%%%%%%%%%
%
%
We have shown that, in the framework of a statistical model,
Bose statistics and AMC together with conservation
of the parity quantum numbers are crucial ingredients in the 
description of the multi-meson channels resulting from $\pqp$ 
annihilation at rest. In such an approach, the projection operator 
formalism appears to be a convenient and consistent method in order
to implement the conservation of non-Abelian quantum numbers.

Isospin conservation alone yields a smooth but much too broad
pion multiplicity distribution, nearly independently of the
temperature $T$, and the interaction volume $V$ is unphysically big.
Only Bose statistics together with conservation of external
quantum numbers provide a reasonable interaction volume of
$V = [5\ldots 20]\ \rm{fm}^3$ in the range of 
$T = [140\ldots 200]\ \rm{MeV}$.
The resulting pion distribution, however, shows no resemblance to a 
Gaussian or Poissonian: either the even or odd numbered $n_{\pi}$ states
are strongly dominating. Thus, even in a statistical approach, once
the effects of quantum number conservation are taken into account,
the appearance of a smooth, Gaussian multiplicity distribution remains a 
mystery. That microscopic dynamical effects from the underlying
QCD mechanism of annihilation should restore the apparent
statistical nature of the empirical multiplicity distribution,
is hard to believe. 

In light of this dilemma, we suggest that the
solution might be buried in the inclusive pion data themselves: the
available bubble chamber data are rather old, and several experiments
with probably different target densities have been exploited.
For this reason, it would be desirable to make a new measurement of the pion
multiplicity distribution at various fixed values of the 
hydrogen density under stable conditions.

\end{document}